\documentclass[amsmath,amssymb,superscriptaddress,letter,11pt]{revtex4}
\usepackage{hyperref}
\usepackage{epsf}
\usepackage{amsfonts,amssymb}
\usepackage{amsmath,amsfonts,times}
\usepackage{graphicx}
\newcommand{\nn}{\nonumber}

\def\p{\partial}

\def\DD{{\sf D}}
\def\Dc{\mathbb C \setminus {\sf D}}

\newcommand{\ti}{\tilde}

\begin{document}
\title{Bubble break-off in  Hele-Shaw flows\\ {\normalsize Singularities and integrable structures}}
\author{Seung-Yeop Lee}
\affiliation{James Frank Institute of the
University of Chicago,   5640 S. Ellis Ave.
Chicago IL 60637}
\author{Eldad Bettelheim}
\affiliation{James Frank Institute of the
University of Chicago,   5640 S. Ellis Ave.
Chicago IL 60637}
\author{Paul Wiegmann}
\affiliation{James Frank Institute of the
University of Chicago,   5640 S. Ellis Ave.
Chicago IL 60637}

\altaffiliation[Also at ]{Landau Institute of Theoretical Physics.}
\date{\today}

\begin{abstract}
Bubbles of inviscid fluid surrounded by a viscous fluid in a
Hele-Shaw cell can merge and break-off.  During the process of
break-off, a thinning neck pinches off to a universal self-similar
singularity.  We describe this process and reveal its integrable
structure: it is a solution of the dispersionless limit of the
AKNS hierarchy. The singular break-off patterns are universal, not
sensitive to details of the process and can be seen
experimentally.  We briefly discuss the dispersive regularization
of the Hele-Shaw problem and the emergence of the Painlev\'e II
equation at the break-off.
\end{abstract}

\maketitle

\section{Introduction}

A Hele-Shaw cell is a narrow gap filled with an incompressible
viscous liquid. When another incompressible fluid with low
viscosity is injected into the cell, two immiscible liquids form a
sharp interface which can be manipulated by pumping in or out the
fluids. The motion of the interface is called Hele-Shaw flow and
it belongs to a large class of processes referred to as Laplacian
growth (LG) \cite{review}.  The latter describes the evolution of
two dimensional domains where the normal velocity of the boundary
is proportional to the {\it conformal measure} --- the gradient of
the conformal map of the domain taken at the boundary. The
dynamics of the interface is unstable; An originally smooth
interface tends to evolve into a complicated and highly curved
finger-like shape consisting of a collection of near-singularities
\cite{Sharon}.

Recent progress in the study of Laplacian growth is due to the
uncovering the integrable structure of the problem and linking to
the universal Whitham hierarchy of soliton theory \cite{M-WWZ,WZ,
KM-WWZ, Shock, Teo2, TW}. The signatures of the integrable
structure of the problem had appeared earlier in the seminal works
\cite{R,Ri,BS,186:Howison,Mineev,Why}. The integrable structure
brings the Laplacian growth into a wider context of mathematical
problems appearing in the theory of nonlinear-waves, and recently
in random matrix theory and string theory. The relations between
the integrable hierarchy and Hele-Shaw flow are transparent and
geometrically illustrative and, therefore, the Hele-Shaw flow may
serve as an intuitive geometric interpretation of the formal
algebraic objects appearing in the theory of non-linear waves and
string  theory (for a recent development see \cite{Shih}). The
relations read:  (i) the real section of the spectral curve---the
central object in Whitham theory---happens to be the interface in
the Hele-Shaw flow, and (ii) the cusp-like singularities of the
interface are linked to shock-waves of dispersive non-linear
waves.

Soliton theory provides a framework for the study and the
classification of singularities emerging in Hele-Shaw flows. A
universal properties close to a separated singularity is described
by a certain reduction of a more general hierarchy.  For instance,
some cusp-like singularities were recently identified with the KdV
hierarchy \cite{Teo2,Shock} and modified-KdV hierarchy \cite{TW},
and have been studied in this framework.

In this paper we study another class  of singularities in the
Hele-Shaw flow: those arising at the {\it break-offs}---the
processes where one bubble with low viscosity breaks into two. We
describe the break-off and show that it is a solution  of
dispersionless limit of the AKNS  hierarchy (dAKNS). Curiously,
the breaking interface has emerged as a target space in the
minimal superstring theory \cite{Shih2}.

As a natural extension of the result, we discuss the dispersive
regularization of the singularities and the emergence of the
Painl\'eve II equation. In the next section, we suggest  an
experimental set-up where the surface tension and other
suppressing mechanisms of singularities are negligible, so that
the universal singular behavior at the break-off can be achieved
experimentally.

\section{Set-up and results}

\begin{figure}
\centering
\includegraphics[width=5cm]{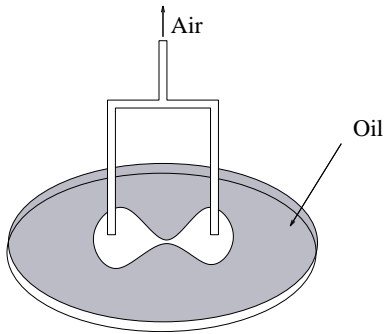}
\caption{A scheme of an experiment where air is drawn from the air
bubble surrounded by oil. A simply connected bubble pinches-off
and eventually breaks-off through a singular neck. After the
break-off both bubbles are maintained at the same pressure. The
singular shape of the neck is universal; It does not depend on the
shape of the original bubble. \label{setup}}
\end{figure}

A Hele-Shaw cell is a narrow gap between two plates filled with
incompressible viscous fluid, say oil. Another immiscible and
relatively inviscid fluid, say air, occupies part of the cell
forming one or several {\it bubbles}.  Most commonly the flow is
produced when air is injected into the cell through a single point
in the cell, pushing oil away to the border of the cell. Here we
will study a different process: air is extracted from the air
bubble---that already exists in the cell---from two fixed points
of the cell situating in the bubble as depicted in
Fig.~\ref{setup}. As the air is drawn out, a simply-connected air
bubble will break into two. The set-up is designed such that the
two separate bubbles after the break-off maintain the same
pressure.  This condition is important.

The  Hele-Shaw flow is driven by  D'Arcy's law: velocity in the
oil domain is proportional to the gradient of the pressure:
\begin{equation}\label{Darcy}
\vec{v} = -\frac{b^2}{12\mu} \vec{\nabla} P.
\end{equation}
The coefficients are the separation of the plates, $b$ and the
viscosity of oil, $\mu$. Below we set the pre-factor
$\frac{b^2}{12\mu}$ to $\frac{1}{2}$, and set the flux-rate such
that the area of the air bubble  is $\pi t$ where $t$ is the time
of the process in the traditional Hele-Shaw flow; $t$ flows
backwards in time in our experimental set-up.  As the oil is
incompressible the pressure is a harmonic function in the oil
domain. Inside the air bubble the pressure is a constant (which
may be set to zero) in time and space. If the surface tension is
negligible then the pressure is continuous across the boundary.
These set up a Dirichlet boundary problem:
\begin{equation}\label{DarcyBoundary}
\Delta P=0 \;\;\mbox{in oil},\quad P=0\;\;\mbox{on the interface},
\quad P\to -\log|z| \;\mbox{ at }z\to\infty,
\end{equation}
where we followed the conventional setup of injection.
$z=x+iy$ is the complex coordinate.

A comment about surface tension is in order.  In the case of an
expanding interface, when the air is injected into the cell,
almost every initial shape of the interface results in a cusp-like
singularity at some finite time, if the surface tension and any
other damping mechanism is neglected. After this time the problem
defined by Eqs.~\ref{Darcy} and \ref{DarcyBoundary} is ill-posed.
Introduction of a damping mechanism (such as surface tension
\cite{Tanveer}) usually ruins the analytical and integrable
structure of the problem. The only exception, where the
singularities are curbed, is the situation proposed in recent
papers \cite{Teo2, Shock}.

However, the reverse process, which we consider in this paper, is
well defined (section 7 of \cite{Shock}).  Before the  break-off
the interface remains smooth regardless of how thin the neck is.
Remarkably the problem remains well defined even after the
break-off when the tips of the two emergent bubbles are highly
curved. This is true when the two bubbles are kept at the same
pressure, which motivates the design of the cell as in
Fig.~\ref{setup}. In this case, the surface tension can be safely
neglected by increasing the extraction rate.  This is perhaps the
only case that the singularities can be regularized without
resorting to a damping mechanism and thus preserving the analytic
structure of the zero surface tension problem.

We will show that the singular behavior of the neck at the
break-off falls into universal classes characterized by two even
integers $(p,q)=(4n,2),\;n=1,2,\cdots$, and  only by a finite
number ($2n$) of parameters.  They are called {\it deformation
parameters}.   While, an infinite number of remaining parameters
characterizing the interface become irrelevant in the vicinity of
the break-off: the break-off is universal.  The deformation
parameters are the {\it flows} of the integrable hierarchy.

  \begin{figure}
    \centering
    \includegraphics[width=10cm]{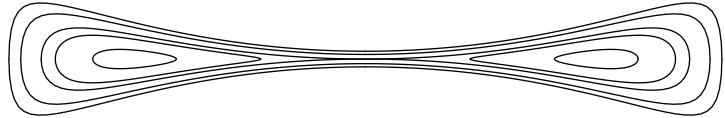}
    \caption{Typical break-off of two droplets at successive time slices.\label{24break-off}}
  \end{figure}

  \begin{figure}
    \centering
    \includegraphics[width=5cm]{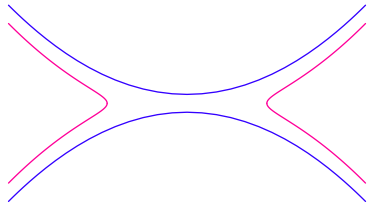}
    \caption{Local curves of (4,2)-break-off just before and after the break-off. \label{24scaled}}
  \end{figure}

In Cartesian coordinates the shape of the neck reads:
\begin{equation}\nn
y^q=x^p F_{p,q}\left(\frac{|t|^{\Delta_{pq}}}{x}\right),
\end{equation}
where we translate the time $t$ such that $t=0$ corresponds to the
break-off point. In this paper we consider only 
$(p,q)=(4n,2)$ with $n$ being a positive integer. We have obtained 
$\Delta_{4n,2}=\frac{1}{2n}$. The function $F_{4n,2}$,
 a polynomial of degree $4n$ at most, is characterized
by $2n$ parameters. Two of these parameters may be eliminated by
re-scaling the coordinates. For example, in the most generic case
of $(4,2)$-break-off, shown in Fig.~\ref{24scaled}, the function
reads
\begin{equation}\nn
F_{4,2}(\xi)
=\begin{cases}\left(1+\frac{1}{2}\xi^{2}\right)^2,
\quad ~t>0:\mbox{before break-off}
\\1-\xi^2~,\quad \quad\quad  t<0:\mbox{after break-off}
\end{cases}.\end{equation}
The same curve will appear as the spectral curve of the AKNS hierarchy.

\section{Global solutions}

Before emphasizing the singular behavior at the break-off, we
start by a general description of Hele-Shaw processes from a
global standpoint. We start from the case of an arbitrary genus
(the genus is the number of bubbles minus one). Then we specify it
to the case of two bubbles using elliptic functions. In this
section we employ a traditional approach of describing the
evolution by a set of moving poles of an analytical function
mapping the exterior of the air bubbles to a standard domain. This
approach has been developed in \cite{R, Ri, BS, Mineev}. It does
not appeal to the notion of integrablilty.

\subsection{The Schwarz function}

Let us consider a bounded, multiply connected domain  $\DD$  in
the complex  plane (a Hele-Shaw cell). The domain is occupied by
an incompressible fluid with negligible viscosity  (air bubbles).
Unbounded exterior of the domain,  $\Dc$, is occupied by an
incompressible fluid with high viscosity (oil). Each air bubble
contains a source through which air can be drawn out of (supplied
into) the bubble such that all bubbles are kept at an equal pressure.
Another neutralizing source is placed at infinity.

We will  assume that the domain $\DD$ is an algebraic or
quadrature domain \cite{Quadratures}, or, more generally, an Abelian
domain \cite{Why}. These domains are defined by the nature of the
Cauchy transform,
\begin{equation}
\label{hh}
h(z)=\frac{1}{\pi}\int_{\Dc}\frac{d^2w}{z-w}
=\begin{cases}h^+(z)\quad \quad\quad \mbox{at}~~ z\in \DD\\
\overline z+h^-(z)\quad  \mbox{at}~~z\in \Dc\end{cases}.
\end{equation}
For quadrature domains the function $h^+$ is rational: containing
a finite number of poles. For Abelian domains $\p h^+$ is rational
and $h^+$ may have logarithmic singularities. The use of Abelian
domains is not very restrictive; Any domain with a smooth boundary
is known to be a limit of Abelian domains \cite{Gus}---the set of
Abelian domain is dense.

The main property of Abelian domains is: the analytical
continuations of the function $\bar z$ from different connected
pieces of the boundary  to the exterior $\Dc$ give the same
result. The function being obtained by this procedure is called a
{\it Schwarz function} \cite{Davis,Shapiro}. In terms of the
Cauchy transform (\ref{hh}),
\begin{equation}\nn
S(z)=h^+(z)-h^-(z).
\end{equation}
If the boundary is smooth, the Schwarz function can be
analytically continued on some strip {\it inside} the domain
$\DD$.

To fix the notations and emphasize the analytical structure of the
Schwarz function in $\Dc$ we write
\begin{equation}\label{Sansatz}
S(z)\sim\sum_{k=1}^K\frac{\mu_k}{(z-q_k)^{h_k}}
+\sum_{l=1}^L\mu_{K+l}\log(z-q_{K+l})
 \end{equation}
where $\sim$ stands for an asymptotic equality near each of the singularities
$q_k\in \Dc$ for $k=1,\cdots,K+L$.  $h_k$'s are positive integers.

The evolution of the domain is determined by the holomorphic function,
\begin{equation}\nn
\phi(z)=\xi(x,y)+ip(x,y),
\end{equation}
where $\xi$ is  {\it stream function} and $p$ is the pressure. In
terms of $\phi(z)$ and the Schwarz function, D'Arcy's law reads,
\begin{equation}\label{EOM1}
\partial_{t}S(z)=i\partial\phi(z).
\end{equation}
An immediate consequence is that the Schwarz function behaves at infinity as,
\begin{equation}\label{last}
\partial_{t}S(z)\sim\overline q_0+\frac{\overline\mu_0}{z}\quad \mbox{at}~~z\rightarrow\infty,\quad q~~\mbox{is constant and}~~\partial_t\overline\mu_0=1.
\end{equation}
which follows form the boundary condition  $\p\phi(z)\to -i/z$ at
infinity. Another consequence is: $h^+$ does not depend on time
(the proof is in Appendix A):
\begin{equation}\nn
\label{h}\partial_{t}h^+(z)=0.
\end{equation}
Therefore the parameters in (\ref{Sansatz}) are fixed and can be
used as the initial data for the  domain.

For a single bubble, these data are sufficient to determine the
evolution.   For a domain consisting of $g+1$ bubbles, additional
$g$ parameters are required \cite{KM-WWZ}; they are, for examples,
relative areas of the bubbles, pressure differences in the
bubbles, or a mixture of the two.  Although later we will focus on
the equal pressure case, here we do not demand this.

\subsection{Dispersionless string equation}

D'Arcy's law written in (\ref{EOM1}) acquires another
important form if the univalent function $\phi$  rather than $z$ is
chosen as a coordinate:
\begin{equation}\label{string1}
\frac{\partial S}{\partial \phi}
\frac{\partial z}{\partial t}-\frac{\partial S}{\partial t}
\frac{\partial z}{\partial \phi}=-i.
\end{equation}
Here the derivative in time is taken at fixed $\phi$. This form is
known as the {\it dispersionless string equation}.

\begin{figure}
\begin{center}
\includegraphics[width=4cm]{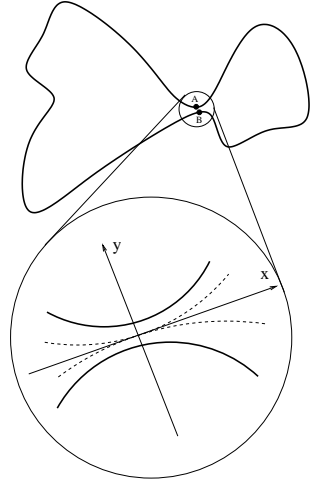}
\caption{Two distant points $A$ and $B$ along the boundary  become close at the break-off.
A blow-up of the critical break-off region is plotted. Dotted lines
are the interfaces when the bubble breaks up.  \label{abouttobreak}}
\end{center}\end{figure}

A Cartesian form of this equation is convenient to study
break-off. Let us choose a coordinate system where the origin is
at the pinch-off point and the $x$ axis is tangent to the
interface at the pinch-off (the dashed lines in
Fig.~\ref{abouttobreak}). It is convenient to use the following
variables.
\begin{equation}\nn
\tilde{x}(z)\equiv\frac{1}{2}(z+S(z)),\quad
 \tilde{y}(z)\equiv\frac{1}{2i}(z-S(z)).
\end{equation}
Note that $\tilde x$ and $\tilde y$ are {\it not}  Cartesian
coordinates of the physical plane; They are holomorphic functions
defined outside the bubble and analytical in some  vicinity of the
boundary.  At the interface, however, $\tilde{x}$ and $\tilde{y}$
{\it are} equal to its Cartesian coordinates. Using the new
variables (with a rescaling of time), Eq.~\ref{string1} is written
as
\begin{equation}\label{string}
\{\tilde y,\tilde x\}\equiv\frac{\partial\tilde y}{\partial \phi}
\frac{\partial\tilde x}{\partial t}-\frac{\partial\tilde y}{\partial t}
\frac{\partial\tilde x}{\partial \phi}=1.
\end{equation}
And equivalently,
\begin{equation}
\partial_{t}\tilde{y}(\tilde{x})
=-\partial_{\tilde{x}}\phi(\tilde{x}).
\label{EOM}
\end{equation}
This form of the evolution equation appeared long ago in
\cite{Kochina}. It was recognized in \cite{M-WWZ} as the
dispersionless string equation known in soliton theory.

\subsection{Hele-Shaw flow as evolving uniformizing map}
It is illustrative to reformulate the Hele-Shaw problem as the
evolution of a  map $z(u)$ uniformizing the exterior domain $\Dc$.
This map maps the exterior domain $\Dc$  consisting of $g+1$
holes, onto a Riemann surface ${\sf U}_+$.

It is customary and convenient to consider  another Riemann
surface ${\sf U}_-$ that is obtained from ${\sf U}_+$  by  an
anti-holomorphic automorphism: $u\to u^*=\overline{f(u)}$ satisfying
\begin{equation}\label{S}
S(z(u))=\overline{z(u^*)}=\overline
z(f(u)).
\end{equation}
At the interface, since $S(z)=\overline{z}$,  we get $u=u^*$: the
interface is invariant under the automorphism. The invariance
allows ${\sf U}_+$ and ${\sf U}_-$ to be glued along the interface
to a Riemann surface without boundary, $\sf U$. The genus of this
surface is $g$---the number of the bubbles minus $1$. This Riemann
surface will be invariant under the {\it square} of the
automorphism, which is in this case called an {\it
anti-holomorphic involution}. The Riemann surface obtained by this
procedure is called  a {\it Schottky double} \cite{Schiffer}; We
shall refer to this Riemann surface as ``the mathematical plane".

Summarizing: $\sf U$ is a Riemann surface with an anti-holomorphic
involution; there are $g+1$ closed cycles (the interfaces) that
are fixed points of the involution; $\sf U$ is divided into ${\sf
U}_+$ (an image of is the oil domain  $\Dc$) and its mirror ${\sf
U}_-$, that border each other along the cycles and are permuted by
the involution.

The  anti-holomorphic automorphism  $u\to u^*$ on the mathematical
plane induces an automorphism on the physical plane:
\begin{equation}\nn
z\to z^*\equiv z(u^*)=\overline{S(z)},
\end{equation}
which is called {\it Schwarz reflection}. The interface is
invariant under the map.

The Advantage of the above construction is that the analytic
functions  defined in the oil domain, such as $S$, $\tilde{x}$,
$\tilde{y}$, and $\phi$, can now be extended to meromorphic
functions on the Riemann surface $\sf U$. From this standpoint a
Hele-Shaw flow is seen as an evolution of a special Riemann
surface. A break-off changes its  genus by one. To describe the
break-off it is sufficient to take two bubbles mapped onto a torus
($g=1$).  Later we will illustrate the above construction on this
example .

We normalize the map such that it  maps a point $u_0$ to infinity.
Being a univalent map it has a simple pole at $u=u_0\in{\sf
U}_+$.
\begin{eqnarray}\label{u0}
z(u)\sim\frac{\alpha_0}{(u-u_0)}
\quad u_0\in {\sf U}_+.
\end{eqnarray}
All the other singularities of $z(u)$ must reside in ${\sf U}_-$.

From (\ref{S}) and (\ref{Sansatz}),  the uniformizing map (of an
Abelian domain) has the following singularity-structure,
compatible with that of the Schwarz function:
\begin{eqnarray}\label{ansatz}
z(u)\sim\sum_{k=1}^K\frac{\alpha_k}{(u-u_k)^{h_k}}
+\sum_{l=1}^L\alpha_{K+l}\log(u-u_{K+l}),
\end{eqnarray}
where $\sum_{k=1}^L\alpha_{K+k}=0$, to ensure the univalency of
$z(u)$. The algebraic nature of the domain warrants the number of
singularities to remain the same during the evolution. Again, they
are all located in ${\sf U}_-$.

In this language the Hele-Shaw flow can be seen as the evolution of
a Riemann surface, the {\it domain} of $z(u)$, which in
its turn is governed by the motion of the singularities: $\alpha_k$'s
and $u_k$'s.  In other words the Hele-Shaw evolution is reduced to
the pole dynamics of the map $z(u)$.

Eq.~\ref{S} relates the singularities of $z(u)$ in (\ref{ansatz})
to the fixed singularities of $S(z)$ in (\ref{Sansatz}). The
relations are explicitly:
\begin{equation}\label{qk}
q_k=z(u^*_k),\quad \mu_k=\overline\alpha_k\left[
\frac{\partial
 z(u^*_k)}{f'(u^*_k)}\right]^{h_k}
\quad k=1,2,\cdots,K+L,
\end{equation}
where we assume $h_k=0$ for $k\geq K+1$. Similarly,
(\ref{last}) and (\ref{u0}) give
\begin{equation}\label{q0}
 q_0=z(u_0^*),\quad \partial_{t}\overline\mu_0=\partial_{t}\left[\alpha_0\,
\overline{\partial z(u^*_0)}{f'(u_0)}\right]=1.
\end{equation}
In the appendix we discuss the derivation of these equations.
Together they give a set of  algebraic equations completely
determining the flow in the case of one bubble.

Additional $g$ constraints will specify either the areas  of the
bubbles or pressure differences. The area of a bubble is the
integral of the Schwarz function along the interface: $\oint_{{\bf
a}_i} S(z) dz$, where ${\bf a}$ is the cycle encircling the
bubble. Pressure-difference between two bubbles, say the $i$th and
$j$th bubble, follows from (\ref{EOM1}).  It is the real part of
the integral of the velocity $\p_{t} S(z)$ along the $b$-cycle
${\bf b}_{ij}$, connecting  $i$th and $j$th bubble.
\begin{eqnarray}\nn
{\rm Im}\,\phi(z)\big|^i_j=
{\rm Re}\int_i^jdz\,\partial_{t}S(z)
=\frac{1}{2}\p_{t}\left(\oint_{{\bf b}_{ij}}S(z)\,dz\right).
\end{eqnarray}
In the following section, we choose the pressures equal, as
required by the set-up in Fig.~\ref{setup}. We would like to
emphasize that the evolution with the same pressures differs
drastically from the evolution with different pressures, which can
also be achieved by controlling the rates of extraction between
two bubbles. We don't consider  such cases here.

\subsection{Two bubbles}

  \begin{figure}
    \centering
    \includegraphics[width=12cm]{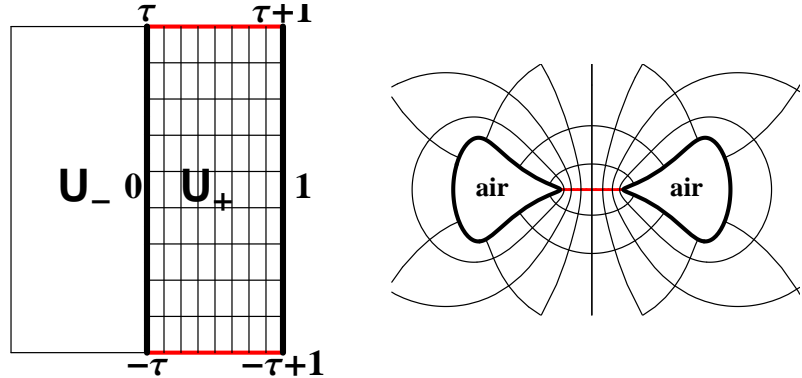}
    \caption{Mathematical plane (left) and the physical domain (right). Conformal map, $z(u)$, maps
    ${\sf U}_+$  to the oil-domain $\Dc$.  We can see that the grid lines in ${\sf U}_+$ are distorted in $\Dc$ by the mapping.  The thick lines at ${\rm Re}(u)=0$ and ${\rm Re}(u)=1$ correspond to the interfaces. The red line at the top (bottom) maps to the critical region. \label{map}}
  \end{figure}

A Riemann surface describing  two bubbles, has the topology of a
torus. The torus is modelled as the complex plane modulo the
periods $2\omega_1$ and $2\omega_2$. The anti-holomorphic
involution is chosen as $f(u)=-u$, and requires $\omega_1$ to be
real and $\omega_2$ to be imaginary \cite{Why}. Choosing a proper
scale  we set $\omega_1=1$ and $\omega_2=\tau$, where $\tau$ is an
imaginary number---the {\it modulus} of the torus---which changes
with time. We take the fundamental domain of $\sf U$ as $-1<{\rm
Re}(u)\leq 1$ and $-|\tau|<{\rm Im}(u)\leq|\tau|$ with the
opposite edges identified as in Fig.~\ref{map}.  The fixed points
of the involution are the lines: ${\rm Re}(u)=0$ and ${\rm
Re}(u)=1$.  Each corresponds to each boundary of the bubbles.

We will follow Refrs.  \cite{Ri,Why}  to construct a map $z(u)$.
For simplicity we discuss only the quadrature domains, where there
is no log-singularity in $z(u)$. The map can be written in terms
of the {\it Weierstrass zeta function} $\zeta(u)$: a
quasi-periodic function satisfying,
\begin{equation}\nn
\zeta(u+2\omega_i)=\zeta(u)+2\eta_i,\quad \eta_i\equiv\zeta(\omega_i)
,\quad i=1,2.
\end{equation}
with a simple pole of residue $1$ at the origin $\zeta(u)\sim
u^{-1}$.

The map $z(u)$  reads
\begin{equation}
\label{map2}
z(u)=\alpha_0\zeta(u-u_0)+\beta+\sum_{j=1}^K\alpha_j\zeta(u-u_j)
\end{equation}
where $u_0\in U_+$ and $u_j\in U_-$ for $j=1,\cdots,K$. The
condition $0=\sum_{j=0}^K\alpha_j$  ensures the periodicity of the
map. It reduces the total number of complex variables (excluding
the modulus $\tau$) to $2K+2$.  The constants of motion are
computed from (\ref{qk}):
\begin{equation}\label{qk1}
\mu_j=\overline{\alpha_j}\partial
z(u^*_j)\quad\mbox{and}\quad q_j=z(u^*_j),\quad (j=1,\cdots,K)
\end{equation}
and from (\ref{q0}):
\begin{equation}\label{q01} q_0=z(u^*_0)\quad\mbox{and}\quad\mu_0=\overline{\alpha_0}\partial z(u^*_0)=t+const.
\end{equation}
Hence the total number of equations is $2K+2$ as well.

The modulus  $\tau$ is chosen to ensure the equal pressure between
the two bubbles:
$0=\frac{1}{2}\partial_t \oint_{\bf b}S(z)\,dz$,
where ${\bf b}$ is a cycle traversing the two bubbles.
Integrating this equation we obtain another constant  of motion, $\Pi$:
\begin{equation}\label{pi}
\Pi=\frac{1}{2}\oint_{\bf b}S(z)\,dz.
\end{equation}
Before a break-off the cycle $\bf b$ does not exist, and at the
break-off it contracts  to a point. Therefore $\Pi$ must be zero
for a break-off: $\Pi=0$, which provides the equation to
determined the modulus. In terms of our conformal map:
\begin{eqnarray}\label{Pi}
0=\Pi=\sum_{j=0}^{K}\eta_1(\alpha_j\overline
  q_j-u_j\overline\mu_j-\overline u_j\mu_j)-\sum_{j,k=0}^{K}\overline\alpha_j\alpha_k\left(
\frac{\wp'(\overline
u_j+u_k)}{2}+\zeta(\overline u_j+u_k)\wp(\overline
u_j+u_k)\right),
\end{eqnarray}
where $\tau$ is implicit in all the elliptic functions.

In the conventional set-up where air is injected into bubbles, a
smooth merging occurs only for a fine-tuned initial condition
which satisfies $\Pi=0$ (\ref{Pi}). If $\Pi\neq 0$, one bubble
develops a singular cusp before the merging. After this moment the
problem without surface tension becomes ill-posed, and requires a
regularization.

\section{Break-off}
To illustrate a break-off we restrict ourselves to a left-right
symmetric case. In this case a bubble will be broken into two
symmetrical bubbles: one is a mirror reflection of the other. The
pressure is also symmetric, and $\Pi$ is vanishing automatically.

It is convenient to follow the reversed evolution, that is, to
consider a merger rather than a break-off.  At the merger the
period ($2\omega_2$) blows up;  a rectangular  fundamental domain
becomes an infinitely long channel---a cylinder; the doubly
periodic function $\zeta(u)$ degenerates into a periodic function
with period $2\omega_1=2$. Accordingly, the (imaginary) modulus
$\tau$ gets infinitely large. Therefore, a convenient small
parameter for an expansion is $e^{i\pi\tau}<1$ (the nome of the
Jacobi theta function) which becomes zero when a merging occurs.
Here we only consider the situation up to the merging point, not
after: $t<0$.

The area near the merger corresponds to the domain near the top
of the rectangle. In $u$-coordinate this region is pushed away to
the infinity as $e^{i\pi\tau}\to 0$. A new coordinate
$\tilde u=u-\tau$ whose origin resides at the top of the rectangle
is proper. We use the expansion in small $e^{i\pi\tau}$:
\begin{equation}\nn
\zeta(u)=\zeta(\tilde u+\tau)
=\frac{\pi^2}{12}(1-24e^{2i\pi\tau})\tilde u+
2\pi e^{i\pi\tau}\,\sin\pi\tilde u+
2\pi e^{2i\pi\tau}\sin2\pi\tilde u+\cdots,
\end{equation}
up to $\tilde u$-independent constants. The map (\ref{map2})
expands:
\begin{eqnarray}
\label{5}
z(\tilde u+\tau)
&=&
C+e^{i\pi\tau}\left(
A_1\sin\pi\tilde u+B_1\cos\pi\tilde u\right)
+e^{2i\pi\tau}\left(
A_2\sin2\pi\tilde u+B_2\cos2\pi\tilde u\right)+\cdots,
\end{eqnarray}
where
\begin{equation}\nn
A_n=2\pi\sum_{k=0}^K\alpha_k\cos n\pi u_k,\quad
B_n=-2\pi\sum_{k=0}^K\alpha_k\sin n\pi u_k,\quad n=1,2.
\end{equation}
For a generic choice of the parameters shape  of the interface is
not physically acceptable: it may contain self-intersecting
interface.

\begin{figure}
\centering
\includegraphics[width=5cm]{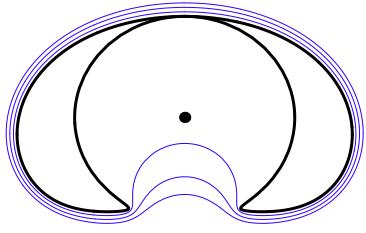}
\caption{Up-down symmetry is broken. Thick line is the boundary.
Thin lines are equi-pressure lines. The dot is one of the
stagnation points. There is a merger in the upper part.  Another
merger in the bottom is imminent, whereupon the bubble will form a
ring-type domain.\label{updown}}
\end{figure}

A particularly simple  configuration occurs by further imposing
up-down symmetry. (A break-off with no up-down symmetry is shown
in Fig.~\ref{updown}.) The full symmetry (up-down and left-right)
may be realized by setting (i) $u_0=\frac{1}{2}$, (ii) the points
$u_k$ in ${\sf U}_-$ are all on the line ${\rm
Re}(u)=-\frac{1}{2}$ symmetrically with respect to the real axis,
(iii) the residues $\alpha_k$'s are all real, and the residues at
poles $u_k$ and $\bar u_k$ are equal,  iv) the merging point is at
the origin: $z(\frac{1}{2} + \tau)=0$. Then we get $C=0$ and
$A_1=B_2=0$.

Within this configuration let us solve the string equation
(\ref{string1}), which is modified into the following form using
(\ref{S}):
\begin{equation}\label{string2}
\partial z(u)\partial_{t}\overline
z(-u)-\partial_{t}z(u)\partial \overline z(-u)
=i\partial\phi(u).
\end{equation}
One use the expansion of $z(u)$ (\ref{5}) to obtain the l.h.s. of
the above equation (\ref{string2}):
\begin{eqnarray}\label{lhs}
(l.h.s.)\sim 2\pi e^{i\pi\tau}\,\frac{d\, e^{2i\pi\tau}} {d t} \cos
\pi \tilde u,
\end{eqnarray}
where the two remaining constants, $A_2$ and $B_1$, are
reduced to 1 by a rescaling of coordinates.

To obtain the r.h.s. of (\ref{string2}) we must find a meromorphic
function $\phi(u)$ that has vanishing imaginary part along the
lines ${\rm Re}(u)=0$ and ${\rm Re}(u)=1$ and has the asymptotic
behavior of $\phi(z)=i\log(u-u_0)\sim-i\log z$ near infinity. One
can construct such function $\phi(z)$ using the Weierstrass sigma
function $\sigma(u)$:
\begin{eqnarray}\nn
\phi(u)=i\log\frac{\sigma(u-u_0)}{\sigma(u-u^*_0)}+2i\,\eta_1{\rm Re}(u_0)\,u.
\end{eqnarray}
The first term renders ${\rm Im}[\phi(u)]$ constant along each
line ${\rm Re}(u)=0$ and ${\rm Re}(u)=1$, then the second term
shifts it to zero for both lines. In the critical regime one can
also expand $\phi(u)$ to produce:
\begin{equation}\nn
(r.h.s.)=i\partial\phi(u)\sim4\pi\, e^{i\pi\tau}\cos\pi\tilde u.
\end{equation}
Combined with the l.h.s. (\ref{lhs}) we obtain $\frac{d\, e^{2i\pi\tau}}{dt}=const.$, or equivalently,
\begin{equation}\nn
e^{i\pi\tau}\sim\sqrt{|t|}
\end{equation}
with $t=0$ being the merging point ($t$ is negative before merging).

The evolution of the interface is described at the leading order:
\begin{equation}\nn
(x,y)\sim(\sqrt{|t|}\ \cosh\pi
s,\;|t|\sinh2\pi s)
\end{equation}
with $s\in{\mathbb R}$ parametrizing the interface.
Alternatively
\begin{equation}\nn
y
=x^2F\left(\frac{\sqrt{|t|}}{x}\right),
\quad F(\xi)=\sqrt{1-\xi^2}.
\end{equation}
This is  the simplest and the most generic (4,2)-class of the
break-off.

\section{Critical regime}
While breaking-off the shape of the interface near the break-off
point does not depend on the details of the rest of the bubble:
locally, the breaking interface is universal. We will call this
regime {\it critical}.

In the critical regime, different scales separate.   It is
convenient to define the scales in the complex plane of
$\phi$---the complex counterpart of pressure---which can be put to
zero at the the tips. As we did in the previous section, it is
convenient to think about break-off in a reversed time. Then the
process appears as a merger.

A critical regime is the image of a small domain containing the
point $\phi=0$. We saw that at $\phi\sim e^{i\pi\tau}$ we  see the
details of the tips of the merging bubbles and details of the neck
after they merge. At $1 \gg \phi \gg e^{i\pi\tau}$, however, the
details of the neck and the tip cannot be seen, and the details
of the remote part of the bubble can not be seen either. At this
scale we see only the asymptote of the neck  $y^q\sim x^p$.
Finally, $\phi = O(1)$ corresponds to the rest of the body of the
bubble. The scale $e^{i\pi\tau}$ changes with time and eventually
disappears at the critical point. In that limit, the shape is
scale invariant as in Fig.~\ref{abouttobreak}. Two even integers
$p$ and $q$ characterize the critical regime. They are sufficient
to determine the scaling functions.

We find solutions only for $(p,q)=(4n,2)$, where $n$ is an integer.
A similar analysis has been carried out for cusp-like singularity
of Hele-Shaw flow in \cite{Teo2}.

\subsection{(4,2)-break-off }

We start from  the most generic (4,2)-class, $\tilde y\propto
A_\pm\tilde x^2$, where the asymptotes may differ at both sides of
the $x$ axis, $A_+\neq -A_-$.  More degenerate $(4n,2)$  will be
addressed later.

After break-off we expect two distinct tips, located respectively
at $x=u_1$ and $x=u_2$.  Assuming these tips are smooth, we have an
expansion: $\tilde x-u_i\propto\tilde y^2+O(\tilde y^3)$, which
leads to branch cut singularities at $u_1$ and $u_2$.
Together with the $(4,2)$ asymptote we are led to the following
form for  the curve:
\begin{equation}
\label{breakoffy} \tilde y(\tilde x)=a(\tilde{x}-b)\sqrt{(\tilde
x-u_1)(\tilde x-u_2)}+c\tilde{x}^2+d\tilde{x}+e,
\end{equation}
with $b$, $u_1$, $u_2$, $d$, and $e$ vanishing at $t=0$.

The potential $\phi(\tilde x)$  must have the same branch points
as the curve, it must behave as $ \ti x$ at large (but not too
large) $\ti x$, and is real on the curve (when $\ti x$ is a real
coordinate of the curve). These requirements determine the
potential:
\begin{equation}
\label{breakoffpressure} \phi = -\sqrt{(\tilde x-u_1)(\tilde
x-u_2)}-\gamma\tilde x.
\end{equation}
The overall normalization can be chosen by a proper rescaling of
time $t$, the coefficient $c$ controls the up-down asymmetry of
the shape shown in the asymptote $\tilde y\propto (\pm a+c)\tilde
x^2$,  the coefficients $e$ and $\gamma$ control the up-down
asymmetry of the velocity. In fact they will be connected by $e=\gamma
t$.

Putting this ansatz into the evolution equation (\ref{EOM}):
$\partial_{t}\tilde y(\tilde x)=-\partial_{\tilde x}\phi(\tilde
x)$, we see that the regular  polynomial part of $\ti y$ and its
singular part consisting of the square root are decoupled. The
solution for the polynomial part set $d=0$, and $c$ as a constant,
and confirms that $e=\gamma t$.

In order to match the singular parts of $\partial_y\tilde y(\tilde
x)$ and $-\partial_{\tilde x}\phi(\tilde x)$, it is sufficient to
match their large $\tilde x$-expansions. $\ti y$ expands as:
\begin{equation}\label{y}
a\Big[\tilde x^2-\Big(b+\frac{u_1+u_2}{2}\Big)\tilde x+
\frac{4(u_1+u_2)b-(u_1-u_2)^2}{8}
+\frac{(2b-u_1-u_2)(u_1-u_2)^2}{16\,\tilde x}+\cdots\Big],
\end{equation}
while  the expansion of the singular part of $-\partial_{\tilde
x}\phi(\tilde x)$ is: $1 +{\cal O}(\ti x^{-2})$. Comparing, we
conclude that the coefficients  of the $\ti x^2,\,\ti x$ and $\ti
x^{-1}$ terms in (\ref{y}) are constants and that the constant
term is proportional to time. The constants must be chosen such
that the shape is  $\ti y=(\pm a+c)\ti x^2$  at time $t=0$. This
is equivalent to setting the parameter $\Pi$ in (\ref{Pi}) to
zero. We obtain $u_1=-u_2=\sqrt{\frac{2|t|}{a}}$ and $b=0$.
Summing up we find an evolving curve (we write it in a
self-similar form):
\begin{equation}
\label{breakoff}  \tilde y=\tilde{x}^2 \left(a\sqrt
{1+\frac{2t}{a\tilde{x}^2}} +c + \gamma \frac{t}{\tilde{x}^2}
\right).\quad\mbox{at}\quad t<0
\end{equation}

Now we turn our attention to the flow after merging (before the
break-off). It is simpler since it doesn't involve any
singularities. When the branch points meet to become a double
point the curve degenerates. The potential  becomes $ \phi(\tilde
x)=\pm\tilde{x}-\gamma \tilde{x}. $ A similar procedure yields the
following curve as the one before the break-off:
\begin{equation}
\label{beforebreakoff} \tilde y=\tilde{x}^2 \left( \pm a \left( 1
+ \frac{t}{a\tilde{x}^2} \right) +c + \gamma
\frac{t}{\tilde{x}^2} \right).\quad\mbox{at}\quad t>0
\end{equation}
Equations (\ref{breakoff}) and (\ref{beforebreakoff}) form our
prediction of the self-similar dynamics for the generic $(4,2)$-break-off.

\subsection{$(4n,2)$-break-offs \label{24nbreakoff}}

There is a larger class of solutions of the Eq.~\ref{EOM} for the
same form of the potential (\ref{breakoffpressure}). They belong
to the $(4n,2)$ class. Below we only consider the curve with
up-down symmetry: symmetric w.r.t. $x$-axis. In this case the
curve and the potential read:
\begin{equation}\label{PP}
\tilde{y}=P(\tilde{x})\sqrt{( \tilde{x}-u_1 ) ( \tilde{x}-u_2) } ,
\quad \phi = -\sqrt{ (\tilde x - u_1) (\tilde x-u_2)},
\end{equation}
where $P(x)$ is a polynomial of degree $2n-1$. Its zeros are
called the {\it double points}. Eq.~\ref{PP} represents a
degenerate hyperelliptic curve, where all but one branch cut shrink to
double points. Double points and the branch points  $u_1$ and $u_2$ move
in time as determined by (\ref{EOM}).

A formally identical problem appeared in the study of
dispersionless integrable hierarchies \cite{kodama, Kono}.
Below we adapt some of the results. The idea is based on the
following observation: the large $\tilde x$-expansion of $\tilde y$
in (\ref{PP}) starts with $2n$ terms of positive powers in $\ti x$, while
the expansion of $\p \phi$ starts from the constant $-1$ and skips the $\ti
x^{-1}$ term. These expansions match only if  the coefficients of
the positive powers of $\tilde{y}$ do not change in time, while
the zeroth order term of the expansion of $\ti y$ is equal to $t$,
and the coefficient in front of $\ti x^{-1}$ is a constant.
Once these terms are matched  the negative powers match automatically.
In the previous section we saw  how it works on the simplest
$(4,2)$ example.

The functions ${H}_k(\tilde x)$, defined below, form a useful
basis of functions with proper analytical behavior: these
functions have the same branch points as $\tilde \phi(\tilde x)$,
behave as $\ti x^k$ at large $\ti x$, and lack all other positive
powers in large $\ti x$-expansion: ${H}_k ( \tilde{x} ) =
\tilde{x}^{k} + O(1)$ for $k=1,2,\cdots$.  These properties
uniquely determine them to be given by the formula:
\begin{eqnarray}
\label{R0H0}
{ H}_k(\tilde x)&=&
\left(\frac{\tilde{x}^{k}}{\sqrt{(\tilde{x}-v)^2-r^2}}\right)_+
\sqrt{(\tilde{x}-v)^2-r^2}
\\\nn&=& \tilde{x}^{k}-\frac{{ R}_{k+1}}{r}-
\frac{{ V}_{k+1}}{\tilde{x}}+\cdots,
\end{eqnarray}
where the subscript $+$ denotes the polynomial part of the large
$\tilde{x}$ expansion. We  set
\begin{equation}\label{44}
v=\frac{u_1+u_2}{2},\quad   r=\frac{u_1-u_2}{2}.
\end{equation}
The two next leading coefficients ${ R}_k  $  and $V_k  $ are
homogeneous polynomials in $r$ and $v$ of degree $k$ which may be
expressed through Legendre polynomials as
\begin{eqnarray}
\label{Legendre} &{
R}_{k+1}=(-i)^k\rho^{k+1}\cosh\varphi\,P_{k}(i\sinh\varphi)
\\\nn
&{ V}_{k+1}=(-i)^{k-1}\rho^{k+1}\frac{\cosh^2\varphi}{k+1}\,
P'_{k}(i\sinh\varphi),
\end{eqnarray}
where
\begin{equation}\nn
\rho^2=r^2-v^2,
\quad ~
\sinh\varphi=\frac{v}{\rho},\quad
\cosh\varphi=\frac{r}{\rho}.
\end{equation}

A general solution is written as a superposition of ${H}_k$
with coefficients $t_k$'s (that will turn out to be time-independent):
\begin{equation}
\label{Sz}
\tilde y(\tilde
x)=\sum_{k=1}^{2n}(k+1)t_{k}{ H}_{k}(\tilde{x}).
\end{equation}
The coefficients are the {\it deformation parameters}. The
differential form of these equation is:
\begin{equation}\nn
\p_{t_k}\ti y=\p_{\ti x}H_{k+1}.
\end{equation}

Matching the leading powers of the expansion (\ref{R0H0}) we get
\begin{align}\nn
&\partial_{t}t_k=0,\quad \mbox{for}\quad k=1,\cdots,2n\\
&t\,r=-\sum_{k=1}^{2n}(k+1)t_{k}{{ R}_{k+1}},\label{hodo3}
\\
&\Pi=-\sum_{k=1}^{2n}(k+1)t_{k}{ V}_{k+1},\nn
\end{align}
where $\Pi$ is a constant of integration. The $2n+2$ conditions
(\ref{hodo3}) determine $2n$ coefficients of the polynomial
(\ref{PP}) and 2 branch points. Thus solving these algebraic
equations is equivalent to solving the Hele-Shaw problem near the
break-off. The last pair of these equations is called the {\it
hodograph relation}.

The constant,  $\Pi$, is the coefficient in front of $\ti x^{-1}$
term of the expansion of $\ti y$, and is exactly the integral
(\ref{pi}) over the ${\bf b}$-cycle of the curve $\tilde y(\tilde
x)$ (\ref{PP})  -  the integral along the cycle encircling the two
branch points. It must vanish for a break-off.

A particularly simple case occurs at $v=0$ when the curve is
left-right symmetric. Then $\Pi=0$, and hodograph equation
(\ref{hodo3}) becomes:
\begin{equation}\nn
t=\sum_{k=1}^n(2k+1)(-1)^{k+1}
\begin{pmatrix}-\frac{1}{2}\\k\end{pmatrix}\,t_{2k}
\,r^{2k},
\end{equation}
which determines the time dependence of $r$.

\subsection{Various break-offs : self-similarity}

Self-similar solutions are of a particular interest. There the
evolution of the curve can be viewed as a scaling transformation
of the coordinate $\tilde y$ and $\tilde x$. They occur when all
the deformation parameter are set to zero except one from even
order which we normalize as
 $t_{2n}=\frac{1}{2n+1}$. As a consequence of the
merging condition $\Pi=0$ we have $v=0$, and $r$ remains as the only scale.

\begin{eqnarray}
\label{(24k)breakoff}
\tilde y(\tilde x)&=&{ H}_{2n}(\tilde
x)
=\left(\frac{\tilde x^{2n}}{\sqrt{\tilde x^2-r^2}}\right)_+
\sqrt{\tilde x^2-r^2},
\end{eqnarray}
The curve enjoys an explicit formula in terms of the Chebyshev
polynomial of the second kind: $U_m( \cosh \theta ) = \frac{
\sinh(m+1) \theta}{ \sinh \theta}$.
\begin{eqnarray}\nn
{ H}_{2k}(\tilde x)&=&2\sinh\theta\,\Big(\frac{r}{2}\Big)^{2k}
\sum_{j=0}^{k-1}
\begin{pmatrix}2k\\j\end{pmatrix}\,
U_{2k-2j-1}(\cosh\theta).\quad\mbox{where}\quad \cosh\theta\equiv\frac{\tilde x}{r}
\end{eqnarray}
The hodograph Eq.~\ref{hodo3} degenerates into $t\,r=\left.-{
R}_{2n+1}\right|_{v=0} =(-1)^{n+1}
\begin{pmatrix}-\frac{1}{2}\\n\end{pmatrix}\,r^{2n+1}
$
and yields
\begin{equation}\nn
r\propto|t|^{1/2n}.
\end{equation}

\begin{figure}
\begin{center}    \includegraphics[width=5cm]{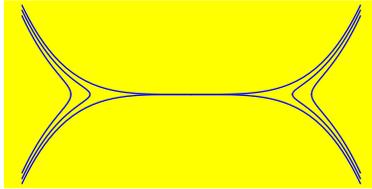}
    \caption{(8,2)-break-off at three successive time slices. Only $t_4$ is turned on.
The curve is $\tilde y=\tilde x(\tilde x^2-\frac{r^2}{2})
\sqrt{\tilde x^2-r^2}$. It shows more singular behavior than the (4,2)-break-off: $r\sim|t|^{1/4}$.
\label{28breakoff}}
 \end{center} \end{figure}

As an example the Fig.~\ref{28breakoff} illustrates ($8,2$)-break-off.

Another interesting case occurs at $u_2\to \infty$ while $u_1$
remains finite. In this case the distance between  the bubbles is
so large that one sees only a tip of one bubble. This case has
been studied in \cite{Teo2}.

\section{Hele-Shaw flows and dispersionless integrable hierarchy}

Here we summarize features of the AKNS hierarchy and its relation to
break-offs in the Hele-Shaw flow. We will show that the flow is
identical to the evolution of the spectral curve appearing in the
dispersionless AKNS (dAKNS) hierarchy. We start by reviewing the
AKNS hierarchy \cite{book, book2}.

\subsection{AKNS hierarchy}

The AKNS hierarchy is an infinite set of non-linear differential
equations written for two functions $r$ and $v$, which depend on
time $t_0$ and an infinite set of deformation  parameters (flows)
$t_k$'s \footnote{In this paper we refer the objects in the
hierarchy by their appearances in the Hele-Shaw flow. It may cause
conflicts with some traditional notations of  soliton theory. For
example the time $t_0$ is often referred to as ``space" in soliton
theory.}. An efficient way of defining the hierarchy requires the
definitions of a few objects, first.

The first is the {\it Lax} operator $L$, a $2\times2$ matrix
consisting of a differential operator $\hbar\p_{t_0}$ (Later we
will consider the limit, $\hbar\to 0$, so we keep $\hbar$ in some of
the formulae in this section), \cite{boost}
\begin{equation}\nn
L=\hbar\partial_{t_0}+U,\quad
U=\frac{1}{2}\left(\begin{array}{cc}
iv & r \\
r & -iv  \end{array}\right).
\end{equation}

The second object is the resolvent $Q(\ti x)$, defined as the
diagonal component of the Green function: $Q(\tilde x) = i\langle
t_0 | (L-i \tilde x \frac{\sigma_3}{2} )^{-1} | t_0 \rangle$,
which is formally expanded in the spectral parameter $\tilde x$ as
\begin{equation}\nn
Q(\ti x)= \sum_{k=1}^{\infty} Q_{k}
\ti x^{-k},\quad \quad \Big(Q_0=i\frac{\sigma_3}{2}\Big)
\end{equation}
where the series of $2\times2$ matrices $Q_k$'s are determined from the commutation relation,
\begin{equation}\label{55}
\left[Q(\tilde x), L-i\tilde x\frac{\sigma_3}{2} \right]=0,
\end{equation}
which, being expanded in powers of $\tilde x$, yield the
recurrence relation for $Q_k$:
\begin{equation}
\label{recurr} \frac{i}{2}[\sigma_3,Q_{k+1}]=\hbar\partial_{t_0} Q_k+[U,Q_k].
\end{equation}
The first few terms are given as
\begin{eqnarray}
\nn
Q_1&=&-r\frac{\sigma_1}{2},
\\\nn
Q_2&=&-rv\frac{\sigma_1}{2}-r'\frac{\sigma_2}{2}
+i\frac{r^2}{2}\frac{\sigma_3}{2},
\\\nn
Q_3
&=&-(\frac{1}{2}r^3+v^2r-r'')\frac{\sigma_1}{2}
-(2r'v+rv')\frac{\sigma_2}{2}
+ir^2v\frac{\sigma_3}{2},
\end{eqnarray}
where ``$~'~$" means the time derivative: $\hbar\p_{t_0}$. For
general $k$, $Q_k$ is a polynomial of degree $k$ in $r$ and $v$,
and their time derivatives.  It follows, from the above definition
of $Q_1$, that the equation $\hbar \p_{t_0} L = [Q_1 - iv\frac{
\sigma_3 }{2}, \, L]$ holds trivially.

The dependence on the deformation parameters can be introduced
accordingly. Including the time in the set of parameters we now
define a hierarchy of flows as \cite{boost}
\begin{equation}
\label{LaxFlow}
\hbar\partial_{t_k}L=[Q_{k+1}-i\hbar\partial_{t_k}\beta\frac{\sigma_3}{2}, L],\quad k=0,1,2,\cdots.
\end{equation}
where $\beta\equiv\frac{1}{\hbar}\int v\,dt_0$. Eq.~\ref{55}
ensures compatibility of the equations: the flows commute.

The first two flow equations are given by
\begin{eqnarray}\label{E}
t_1&:&
\pm i\hbar\partial_{t_1}(re^{\pm i\beta})=(re^{\pm i\beta})''-\frac{r^2}{2}(re^{\pm i\beta})
\\\nn
t_2&:&\begin{cases}
\hbar\partial_{t_2}r=\frac{3}{2}r^2r'+3r'v^2+3rvv'-r'''\\
\hbar\partial_{t_2}\beta=\frac{3}{2}r^2v+v^3-3\frac{r'}{r}v'-3\frac{r''}{r}v-v''
\end{cases}
\end{eqnarray}

The recursion relations (\ref{recurr}) and the flow equations
(\ref{LaxFlow}) are more explicitly written in terms of the
components of $Q_k$,
\begin{equation}\nn
Q_k\equiv-{\cal R}_k\frac{\sigma_1}{2}+\Theta_k\frac{\sigma_2}{2}+i{\cal V}_k\frac{\sigma_3}{2}.
\end{equation}
They are: $r\Theta_k=-{\cal V}'_k$, and
\begin{eqnarray}\label{EE}
&{\cal R}_{k+1}=r{\cal V}_k+v{\cal
R}_k-\left(\frac{{\cal V}'_k}{r}\right)',\\\label{middle}
&{\cal V}'_{k+1}=v{\cal V}'_k+r{\cal R}'_k,
\end{eqnarray}
and the flow equations (\ref{LaxFlow}):
\begin{eqnarray}
\partial_{t_k} r=\frac{{\cal V}'_{k+2}}{r},\quad   \label{hierarchy}
\partial_{t_k} v=\left(\frac{{\cal R}_{k+2}}{r}\right)'.
\end{eqnarray}
It follows  that the flows can be written in an integrated form  \cite{KMS}
\begin{eqnarray}\label{hodoAKNS}
0&=&\sum_{k=0}(k+1)t_{k}{\cal R}_{k+1}
\\\nn -\Pi&=&\sum_{k=0}(k+1)t_{k}{\cal V}_{k+1}
\end{eqnarray}
with a constant $\Pi$.  Alternatively, this hodograph-type
relations (\ref{hodoAKNS})  may be used as a definition of the
hierarchy. The reader may notice that the structure of these
equations is similar to the hodograph equations in (\ref{hodo3}).

There are several known reductions of AKNS hierarchy.

\begin{itemize}
\item [-] NLS: When $r$ is purely imaginary, the first flow
equation of the hierarchy is non-linear Schroedinger equation.
\begin{equation}\nn
 0=i\dot\psi+\psi''+\frac{|\psi|^2}{2}\psi,\quad \psi=\pm ire^{-i\beta},\quad
(\mbox{NLS equation})
\end{equation}
This reduction seems to be irrelevant for the Hele-Shaw problem.
Other reductions are relevant. \item[-] mKdV: When $v=0$ and
$\kappa=\pm i r$. The even flows constitute mKdV hierarchy. The
first equation of the hierarchy is the $t_2$-flow:
\begin{equation}\nn
0=\dot \kappa+\frac{3}{2}\kappa^2\kappa'+\kappa'''.\quad \quad  ~(\mbox{mKdV equation})
\end{equation}
where ``~$\dot{}$~" stands for $\hbar\p_{t_2}$.

\item[-] N-AKNS: It occurs if all the deformation parameters
vanish except: $t_N=\frac{1}{N+1}$. The first interesting case
occurs at $N=2$. In this case (\ref{hodoAKNS}) reads
\begin{eqnarray}\nn
&\frac{r^3}{2}+v^2r-r''+t_0 r=0
\\\nn&
r^2v+\Pi=0
\end{eqnarray}
which, after elimination of $v$ yields the (modified) Painlev\'e
II equation
\begin{equation}\label{P2}
r''=\frac{r^3}{2}+\frac{\Pi^2}{r^3}+t_0 r.
\end{equation}
\end{itemize}

\subsection{Linear problem and $M$ operator}
The flow equations appear as a result of compatibility of the spectral problems:
\begin{equation}\label{Lax}L\Psi=i\tilde{x}\frac{\sigma_3}{2}\Psi,
\end{equation}
and,
\begin{equation}\label{flows}
\hbar\partial_{t_k} \Psi=\Big({\cal H}_{k+1}(\tilde x)-i\hbar\partial_{t_k}\beta\frac{\sigma_3}{2}\Big)\Psi,
\end{equation}
where
${\cal H}_{k}(\tilde x)=
Q_0\tilde x^k+Q_1\tilde x^{k-1}+\cdots+Q_k$. \cite{boost}

Another important object of the hierarchy is  the  $M$-operator
defined as
\begin{equation}\label{M1}
M(\tilde x)\Psi=\partial_{\tilde x}\Psi.
\end{equation}
Its relation to flows follows from compatibility of the above definition with the Lax equations (\ref{Lax},\ref{flows})
 \begin{equation}
\label{M}
M(\tilde x)=\sum (k+1)t_k
\Big({\cal H}_{k}(\tilde x)-i\hbar\partial_{t_{k-1}}\beta\frac{\sigma_3}{2}\Big).
\end{equation}
The proof is standard in the theory of integrable hierarchy. We
sketch it in the Appendix D.

The relation expressing the compatibility of (\ref{Lax}) and (\ref{M1}):
\begin{equation}
\label{Mstring}
[M(\tilde x),L-i\tilde x\frac{\sigma_3}{2}]=-i\frac{\sigma_3}{2},
\end{equation}
is sometimes referred to as the string equation. The $M$-operator
reflects homogeneous properties of the wave function $\Psi$ with
respect to the scaling transformations:
$t_k\to\lambda^{k+1} t_k$, and $(\ti x,r,v)\to (\frac{\ti x}{\lambda},\frac{r}{\lambda},\frac{v}{\lambda})$.

\subsection{Dispersionless limit and genus-0 Whitham equations}
The Hele-Shaw problem appears as a dispersionless limit of the
hierarchy as $\hbar\to 0$.

Let us  pursue the limit on a formal level. As is typical for a
semiclassical approximations  the wave function has the form
\begin{equation}\label{psi}
\Psi\sim e^{-\frac{1}{\hbar}{\cal A}},
\end{equation}
where the ``momentum,"  $\phi=2i\p_{t_0}{\cal A}$, obeys the spectral equation
\begin{equation}\nn\det \left(\frac{i}{2}\phi+U-i\tilde x\frac{\sigma_3}{2}\right)=0.
\end{equation}
The latter gives a  relation between the  ``momentum"
$\phi$ and the spectral parameter
\begin{equation}\nn
\phi^2(\ti x) = (\tilde x-v)^2-r^2,
\end{equation}
and determines a spectral surface---a Riemann surface where the
wave function is univalent.

Accordingly, the commutator in (\ref{Mstring}) is replaced by a
Poisson bracket. If $\ti y(\ti x)$ is an eigenvalue of the
operator $M(\ti x)$, treated as a function on the spectral
surface, the pair $\ti y(\ti x(\phi))$ and $\ti x(\phi)$ obey the
dispersionless string equation
\begin{equation}\nn
\{\tilde y,\tilde x\}=1.
\end{equation}
An alternative form of this equation follows from (\ref{M1}),
\begin{equation}\label{S1}
\p_{t_0} y=-\p_{\ti x}\phi
\end{equation}
where the time derivative is taken at a fixed $\ti x$.

The dispersionless limit of Eqs.~\ref{E},\ref{EE},\ref{P2} is
easy. One simply drops the time derivative there.  In this limit
the objects of AKNS hierarchy converge to the objects discussed in
the previous section \ref{24nbreakoff}.
\begin{eqnarray}\nn
&{\cal R}_k\to R_k,\\\nn
&{\cal V}_k\to V_k,\\\nn
&\sqrt{\det|{\cal H}_k-i\frac{{\cal R}_{k+1}}{r}\frac{\sigma_3}{2}|}\to H_k.
\end{eqnarray}
Under this identification (\ref{hodoAKNS}) become (\ref{hodo3}),
and the Hele-Shaw dynamics is seen to be governed by the
dispersionless AKNS hierarchy.  The dispersionless limit can be
also understood as genus-0 Whitham equations.  We briefly discuss
it below. For a comprehensive description of Whitham theory see
\cite{Whitham}.

Integrable equations have special solutions known as finite-gap
solutions. They  are quasi-periodic in times. For these cases, the
wave-function $\Psi$ as a function of the spectral parameter $\ti
x$ is multi-valued. It is a univalent on a spectral surface. The
case when $r$ and $v$ do not depend on time $t_0$ is the simplest
example. The period of this solution is infinite, but the wave
function of this solution oscillates with a period $\sim
\hbar^{-1}$. Then  the formula (\ref{psi}) gives an exact
solution.  The genus of the spectral surface is zero.

A class of non-periodic solutions may be obtained from periodic
solutions under assumption that  periods (the moduli) of the
solution depend on time but change  slowly, such that they do not
change much at the time of the order to $\hbar$. Constructions of
slowly modulated solutions is the subject of Whitham theory.

Modulation of genus-0 solutions is simplest. In order to obtain it
one simply drops the time derivatives in most equations as it has
been discussed above.  By doing so we obtain the same equations as
obtained for the break-off in the previous section.

\subsection{Hele-Shaw flow and dispersionless limit of integrable hierarchy}

As we have seen, the dispersionless limit of AKNS hierarchy and
equations for the Hele-Shaw flow close to a break-off/merging are
identical. This is not an accident. In \cite{KM-WWZ} an
equivalence  between Hele-Shaw and the universal Whitham hierarchy
has been established on general grounds.

The relation between integrable hierarchies and Hele-Shaw flows
can be summarized as follows: in the limit $\hbar\to 0$
\begin{itemize}
\item[-] the eigenvalues of the Lax operator become different
branches of the potential $L\to\phi(\ti x)$; \item [-] the
operator $M$ becomes the height function $M\to \ti y(\ti x)$;
\item[-]  the real section of the spectral curve $\ti y(\ti x)$
when $\ti y=y,\,\ti x=x$ is identified with the form of the
interface; \item[-] Eq.~\ref{Mstring} converts to D'Arcy's law
written in the form (\ref{string}); \item[-] The integrable
hierarchy describe the response of the interface to a variation of
deformation parameters.
\end{itemize}

\section {Hele-Shaw problem as a singular limit of  dispersive
non-linear waves} A relation  between the Hele-Shaw problem  and
the dispersionless limit of integrable equations  reveals the
nature of fingering instabilities and singularities of the
Hele-Shaw problem. It is known for a long time that the
dispersionless limit $\hbar\to 0$ of dispersive non-linear waves
is singular \cite{singular}. Despite the fact that the integrable
equations are well defined at all times, the $\hbar\to 0$ limit of
some---the most relevant---solutions cannot be taken after some
finite time of evolution. These solutions develop singular
features at small scales of the order $\hbar$ and have no meaning
at $\hbar=0$. These singularities appear in non-linear waves as
shock waves. In Hele-Shaw flows where the surface tension is
omitted they appear as  cusp-like singularities.

Some solutions, however, do not lead to a singularity. One of them
has been considered in this paper. An inevitable break-off of the
extracting process in Fig.~\ref{24break-off}, although singular,
is well defined at all time even in the limit of zero-surface
tension.

However, an injection process, unless it is fine-tuned to satisfy
$\Pi=0$,  leads to cusp-singularities.

\begin{figure}
\begin{center}    \includegraphics[width=6cm]{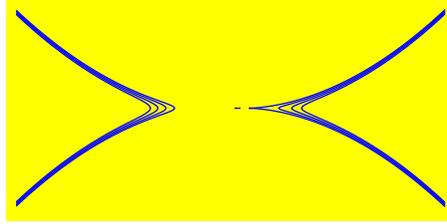}
    \caption{(4,2)-break-off with nonzero $\Pi$. One of the tip develops a (2,3)-cusp before merging.
   Evolution beyond this point requires a regularization of the cusp-singularity.
\label{nonzeropi}}
 \end{center} \end{figure}

\begin{figure}
\begin{center}    \includegraphics[width=6cm]{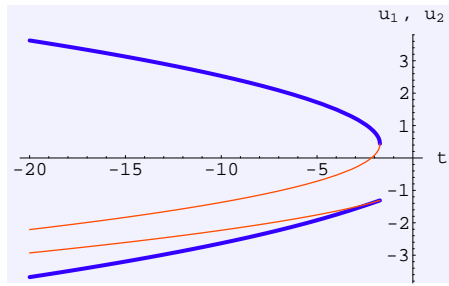}
\caption{Trajectory of the branch points $u_1$ and $u_2$ from
solving the dispersionless limit of Painlev\'e II equation
(\ref{82}).  Blue (thick) line is the physical solution and the
red (thin) line is the unphysical solution.  Being continued
through the cusp, the solution becomes multi-valued and therefore
unphysical. \label{overhang}}
 \end{center} \end{figure}

We can summarize this on a simple example of Painl\'eve II
Eq.~\ref{P2}
\begin{equation}\label{P22}
\hbar^2 \p_t^2r=\frac{r^3}{2}+\frac{\Pi^2}{r^3}+t_0 r.
\end{equation}
As we have mentioned the $\hbar\to 0$ limit of this equation,
\begin{equation}\label{82}
0=\frac{r^3}{2}+\frac{\Pi^2}{r^3}+t_0 r,
\end{equation}
describes  the critical regime  when all deformation parameters
$t_k$ vanish for $k>2$. In this limit  $r$ and $v$  have a simple
interpretation: $r$ is a the distance between two tips,
$v=-\frac{\Pi}{r^2}$ is a measure of asymmetry of the bubbles
(Fig.~\ref{nonzeropi}).  When $\Pi=0$ the bubbles merge according
to the law $r=\sqrt{-2t_0}$.  When $\Pi\neq 0$, however,
Eq.~\ref{82} happens to have four real solutions. The graph of
$r(t_0)$ overhangs at some time $t^*=-\frac{3\Pi^{2/3}}{2^{4/3}}$
with $r(t^*)=2^{1/3}\Pi^{1/3}$ as in Fig.~\ref{overhang}. At this
point the tip of one droplet develops a (2,3) cusp singularity
$y^2\sim x^3$ as shown in Fig.~\ref{nonzeropi}. Equation
(\ref{82}), as well as the original un-regularized D'Arcy law
(\ref{EOM}), have no sense beyond this point.

However, the Painlev\'e II Eq.~\ref{P22} is well defined at any
time.  The approximation of simply dropping the dispersion term
$\hbar^2 \p_t^2r$ is invalid. The reason is simple: an initially
smooth  function such as $r(t)$ will get sharper, such that the
dispersion is no longer small. In fact the approximation becomes
invalid some time before the overhang occurs. This phenomena is
known in the theory of non-linear wave as {\it a gradient
catastrophe}.

The relation between  the integrable equations and the Hele-Shaw
problem raises a natural question. Can the Hele-Shaw problem be
extended beyond $\hbar\to 0$ limit? Can the parameter $\hbar$ be
used to regularize singularities in a meaningful physical way?
These questions will be the subject of future study.

\section{Conclusion}

We have analyzed the universal shapes of bubbles in a Hele-Shaw
cell at break-off. These are given by the Eq.~\ref{breakoff}. At
the pinch-off point the shapes are universal and self-similar:
their evolution is reduced to a proper re-scaling of the
coordinates.

The universal shapes may be observed in a proposed experiment
under the condition that, after break-off, the  bubbles are all
maintained at the same pressure.

We have extended the relation between Hele-Shaw flow and integrable
hierarchies to the case of a break-off processes. We have shown
that the universal shapes are solutions to the dispersionless
limit of the AKNS hierarchy and Painlev\'e II equation.
\paragraph*{Acknowledgments}
We thank L. Kadanoff for discussion and interest to this work. The
work was supported by the NSF MRSEC Program under DMR-0213745, NSF
DMR-0220198. E.B. acknowledges H. Swinney for discussions.

\begin{appendix}

\section{Richardson's theorem}
Here we prove that $h^+(z)$ does not change in time. The evolution
of the domain $\DD\to\DD+\delta\DD$ means an evolution of the
function as follows:
\begin{equation}\label{trap}
\delta h^+(z)= -\frac{1}{\pi} \int_{\delta D} \frac{d^2w} {z-w}
\end{equation}
where $\delta \DD$ is the area trapped between the two contours
$\partial \DD$ and $\partial( \DD+\delta\DD)$. Its area  is given
by D'Arcy's law $\delta t\,|\vec v\times\vec{dl}| = \frac{1}{2}
\partial \phi(z) \, dz$, where $\vec{dl}$ is a tangent line element
to the contour $\partial D$. This transforms the variance of the
function (\ref{trap})  to an integral holomorphic function
\begin{equation}\nn
\delta h^+(z)=-\frac{1}{2\pi}\oint_{\partial D}\frac{1}{z-w}
\partial_w\phi(w)\,dw=0
\end{equation}
which vanishes since the integrand is an analytical function in the
exterior domain.

\section{Singularities of the Schwarz function}
Now we examine the singularities in $S(z)$ (\ref{qk},\ref{q0}).

Let us first pick a pole singularity $u_k$ . In its vicinity
$S(z)$ diverges as
\begin{eqnarray}\nn S(z)&=&\overline{z(u^*)}\sim
\frac{\overline\alpha_k}{(\overline{u^*-u_k})^{h_k}}
=\frac{\overline\alpha_k}{(f(u)-f(u^*_k))^{h_k}}
\\\nn&\sim&\frac{\overline\alpha_k}{f'(u^*_k)^{h_k}}
\frac{1}{(u-u^*_k)^{h_k}}
\end{eqnarray}
where we used $u=\overline{f(u^*)}$.
In its turn
$z(u)-z(u^*_k)\sim z'(u^*_k)(u-u^*_k)$ as $u^*_k$ locates in ${\sf U}^+$, where $z(u)$ is analytical.
Therefore the Schwarz function
\begin{equation}\label{S11} S(z)\sim\overline\alpha_k\left[
\frac{z'(u^*_k)}{f'(u^*_k)}\right]^{h_k}
\frac{1}{(z-z^*_k)^{h_k}}\quad \quad 1\leq k\leq K
\end{equation}
is singular at
$z^*_k\equiv z(u^*_k)$.
Similarly, for the logarithmic singularities yield
$\label{S2} S(z)\sim\overline\alpha_k
\log(z-z^*_k)
$.

Finally,  the singularity (\ref{u0}) at $u_0$ determines the
behavior of $S(z)$ at $z\to\infty$
\begin{eqnarray}\nn S(z)&=&\overline{z(u^*)}\sim
\overline{z(u^*_0)}+\overline{\partial z(u^*_0)(u^*-u^*_0)}
\\&\sim& \overline{z(u^*_0)}+\overline{\partial
z(u^*_0)}{f'(u_0)}({u-u_0})\sim \overline{z^*_0}+\alpha_0\,
\overline{\partial z(u^*_0)}{f'(u_0)}\,\,\frac{1}{z}.
\end{eqnarray}

\section{Boost symmetry}

Conventionally, in the AKNS system, the matrix $U$ is defined to
be purely off-diagonal \cite{book,book2}. It requires some
similarity transformation, called ``boost" symmetry \cite{KMS}, to obtain our form. The ``un-boosted" Lax operator is
written as
\begin{eqnarray}\nn
\tilde L=e^{\frac{i}{2}\beta\sigma_3}L \,e^{-\frac{i}{2}\beta\sigma_3}
=\hbar\partial_{t_0}+\frac{r}{2}\left(\begin{array}{cc}
0 & e^{i\beta} \\
e^{-i\beta} & 0  \end{array}\right),
\end{eqnarray}
where we have defined $\beta\equiv\frac{1}{\hbar}\int v\,d{t_0}$.

One may also define the un-boosted $Q_k$ as $\tilde Q_k = e^{
\frac{i}{2} \beta \sigma_3} Q_k \,e^{ -\frac{i}{2} \beta \sigma_3
}$. The flow Eq.~\ref{LaxFlow} are more natural in this frame.
\begin{equation}
\nn
\hbar\partial_{t_k}\tilde L=[\tilde Q_{k+1},\tilde L],\quad k=0,1,2,\cdots,
\end{equation}
Also Eq.~\ref{flows} takes a simpler form in this frame:
$$\hbar\partial_{t_k}(e^{\frac{i}{2}\beta\sigma_3}\Psi)=\tilde{\cal
H}_{k+1}(\tilde x)\,(e^{\frac{i}{2}\beta\sigma_3}\Psi)$$ with
$\tilde{\cal H}_k=\tilde Q_0\tilde x^k+\tilde Q_1\tilde
x^{k-1}+\cdots+\tilde Q_k$.

\section{$M$-operator and the flows}
Here we derive the formula
\begin{equation}
\nn
M=\sum_{k=1}^\infty (k+1)t_k \partial_{t_{k-1}},
\end{equation}
from the string Eq.~ \ref{Mstring}. It is then obvious that, by
acting $M$ on the wave-function $\Psi$ using Eq.~\ref{flows}, one
gets Eq.~\ref{M}. Here, instead, we are going to prove by acting
$M$ on the un-boosted wavefunction,
$\tilde\Psi=e^{i\beta\frac{\sigma_3}{2}}\Psi$.

We first write the  $M$-operator as a linear sum of all the flows
$M(\tilde x)=\sum_{k=1}^\infty c_k\,\partial_{t_{k-1}}$
and prove the coefficients $c_k$ to be $(k+1)t_k$.  The relation, $[M-\partial_{\tilde x},\partial_{t_k}-\tilde{\cal H}_{k+1}]=0$, by acting on $\tilde\Psi=e^{i\beta\frac{\sigma_3}{2}}\Psi$, gives the relation
\begin{eqnarray}
\nn0&=&\partial_{\tilde x}\tilde{\cal H}_{k+1}-\partial_{t_k}\big(\sum_{j=1}^\infty c_j\tilde{\cal H}_j\big) M+[\tilde{\cal H}_{k+1},\big(\sum_{j=1}^\infty c_k\tilde{\cal H}_j\big)]
\\\nn&=&
=\partial_{\tilde x}\tilde{\cal H}_{k+1}-\sum_{j=1}^{\infty}
\left[c_j\left(
\partial_{t_k}\tilde{\cal H}_j-[\tilde{\cal H}_{k+1},\tilde{\cal H}_j]\right)+(\partial_{t_k}c_j)\tilde{\cal H}_j
\right]
\\\nn&=&\partial_{\tilde x}\tilde{\cal H}_{k+1}-\sum_{j=1}^{\infty}
\left[c_j\,\partial_{j-1}\tilde{\cal H}_{k+1}+(\partial_{t_k}c_j)
\tilde{\cal H}_j\right]
=\partial_{\tilde x}\tilde{\cal H}_{k+1}-\left(\partial_{\tilde x}\tilde{\cal H}_{k+1}
-(k+1)\tilde{\cal H}_k\right)
-\sum_{j=1}(\partial_{t_k}c_j)\tilde{\cal H}_j
\\\nn&=&(k+1)\tilde{\cal H}_k-\sum_{j=1}(\partial_{t_k}c_j)\tilde{\cal H}_j,
\end{eqnarray}
where we have used $ \sum_{j=1}
c_j\,\partial_{j-1}\tilde{\cal H}_{k+1}=\partial_{\tilde x}\tilde{\cal H}_{k+1}-(k+1)\tilde{\cal H}_k, $
following from $ [R(\tilde x),\partial_{\tilde x}-M]=0. $ This
gives $c_k=(k+1)t_k$.

\end{appendix}


\begin{thebibliography}{77}

\bibitem{review}
For a review see D. Bensimon et al., {\it Viscous flows in two dimensions}, Rev. Mod. Phys. {\bf 58} 977-999
(1986); B. Gustafsson and A. Vasil'ev, http://www.math.kth.se.

\bibitem{Sharon} E. Sharon, M. G. Moore,
W. D. McCormick, and H. L. Swinney, {\it Coarsening of fractal viscous fingering patterns}, Phys. Rev. Lett. {\bf 91}, 205504 (2003).

\bibitem{M-WWZ}
M. Mineev-Weinstein, P. B. Wiegmann and A. Zabrodin, {\it Integrable structure of integrable dynamics}, Phys. Rev.
Lett. {\bf 84} 5106-5109 (2000).

\bibitem{WZ}
P.B. Wiegmann and A. Zabrodin, Commun. Math. Phys. {\bf 213}
523-538 (2000).

\bibitem{KM-WWZ}
I. Krichever, M. Mineev-Weinstein, P. Wiegmann and A. Zabrodin,
Physica D {\bf 198} 1-28 (2004).

\bibitem{Shock} E. Bettelheim, P. Wiegmann, O. Agam and A. Zabrodin,
  {\it Singular limit of Hele-Shaw flow and dispersive regularization
  of shock waves}, Phys. Rev. Lett. {\bf 95}, 244504 (2005).

\bibitem{Teo2} R. Teodorescu, A. Zabrodin and P.B. Wiegmann, {\it Hele-Shaw problem and the KdV hierarchy}, Phys. Rev. Lett. {\bf 95} 044502 (2005).

\bibitem{TW} R. Teodorescu, and P.B. Wiegmann, unpublished.

\bibitem{R}
S. Richardson, J.Fluid Mech. {\bf 56} 609 (1972)

\bibitem{Ri} S. Richardson, Journal of Fluid Mechanics {\bf 56}, 609
  (1972); Eur. J. App. Math. {\bf 12}, 571 (2001);{\bf 12}, 665 (2001).

\bibitem{BS} B. Shraiman and D. Bensimon, Phys. Rev. A. {\bf 30}, 2840 (1984)

\bibitem{186:Howison}
S.~D. Howison, Journal of Fluid Mechanics {\bf 167}, 439 (1986).

\bibitem{Mineev} M. Mineev-Weinstein, Phys. Rev. Lett. {\bf 80}, 2113 (1998).

\bibitem{Why}
P. Etingof and A. Varchenko, {\it Why does the boundary of a round
  drop becomes a curve of order four}, University Lecture Series,
vol. 3, American Mathematical Society, Providence, RI, 1992

\bibitem{Shih} N. Seiberg and D. Shih, {\it Minimal String Theory}, eprint  hep-th/0409306.

\bibitem{Shih2} N. Seiberg and D. Shih, {\it Flux Vacua and Branes of the Minimal Superstring}, JHEP 0501 (2005) 055

\bibitem{Tanveer} S. Tanveer, Phil. Trans. R. Soc. A {\bf 343} 155 (1993).

\bibitem{Quadratures} D. Aharonov and H. S. Shapiro, J. Anal.
Math. {\bf 30} (1976), 39; B. Gustafsson, Acta Appl. Math. {\bf 1}
(1983), 209; D. Crowdy and H. Kang , Journal of NonLinear Science,
{\bf 11} (2001), 279

\bibitem{Gus}
B. Gustafsson, Acta Appl. Math. 1 (1983) 209-240.

\bibitem{Davis}
P. J. Davis, The {S}chwartz function and its applications, {The
Mathematical Association of {A}merica}, {USA}, {(1974)}

\bibitem{Shapiro}
H. Shapiro, The Schwarz function and its generalization to higher dimensions, University of Arkansas Lecture Notes in the Mathematical Sciences, Volume 9, W.H. Summers, Series Editor, A Wiley-Interscience Publication, John Wiley and Sons, 1992.

\bibitem{Kochina}
P. Ya. Polubarinova-Kochina, Cokl, Akad. Nauk SSSR {\bf 47} 254
(1945); P. P. Kufarev, ibid {\bf 57} 335 (1947).

\bibitem{Schiffer}
M. Schiffer, D.C. Spencer, {\it Functionals of Finite Riemann
  Surfaces}, Princeton University Press, 1954.

\bibitem{kodama} Y. Kodama, and B. G. Konopelchenko, {\it Deformations of plane algebraic curves and integrable systems of hydrodynamics type}, eprint arXiv:nlin.SI/0210002, Oct 2002.

\bibitem{Kono} B. Konopelchenko and L. Mart\'{i}nez Alonso, {\it Integrable quasiclassical deformations of algebraic curves}, J. Phys. A:Math. Gen. {\bf 37} (2004) 7859-7877.

\bibitem{book} L. A. Dickey, {\it Soliton Equations and Hamiltonian Systems},
 Advanced series in mathematical physcis; v.26 (2003) World Scientific, Singapore.

\bibitem{book2}  F. Gesztesy, and H. Holden, {\it Soliton Equations and Their Algebro-Geometric Solutions}, Cambridge Studies in Advanced Mathematics; Vol. 79. (2003) Cambridge University Press, Cambridge, 2003.

\bibitem{KMS} I.R. Klebanov, J. Maldacena, and N. Seiberg, {\it
  Unitary and Complex Matrix Models and 1-d Type 0 Strings}, Commun. Math. Phys. {\bf 252} (2004) 275-323, hep-th/0309168.

\bibitem{Whitham} Whitham, G. B., Nonlinear Dispersive Waves, SIAM Journal Appl. Math, {\bf 14} (4). 956-958, 1966.

\bibitem{Krichever} I. Krichever, Funct. Anal. Prilozh., 22 (1988), 200-213.

\bibitem{GP} Gurevich, A. V. and Pitaevski, L. P., Sov. Phys. JETP, 38 (2), 291-297, 1974

\bibitem{singular} Singular Limit of Dispersive Waves, eds. N.M. Ercolani et.al., Plenum Press, New York (1994).

\bibitem{boost} See the Appendix. C.

\end{thebibliography}
\end{document}